\documentstyle[aasms4,12pt,epsf]{article}

\begin{document}

\title{The Local Luminosity Function at 25 Microns}

\author{David L. Shupe, Fan Fang, Perry B. Hacking}
\affil{Infrared Processing and Analysis Center, Jet Propulsion Laboratory, 
Caltech 100-22, Pasadena, CA 91125}

\and

\author{John P. Huchra}
\affil{Center for Astrophysics, 60 Garden Street, Cambridge, MA 02138}

\begin{abstract}

The local luminosity function at 25 $\mu$m provides the basis for
interpreting the results of deep mid-infrared surveys planned or in
progress with space astrophysics missions including ISO, WIRE and SIRTF.
 We have selected a sample of 1458 galaxies from the IRAS Faint Source
Survey with a flux density limit of 250 mJy at 25 $\mu$m.  The local
luminosity function is derived using both parametric and non-parametric
maximum-likelihood techniques, and the classical $1/V_{max}$ estimator. 
Comparison of these results shows that the $1/V_{max}$ estimate of the
luminosity function is significantly affected by the Local Supercluster.

A maximum-likelihood fit to the radial density shows no systematic
increase that would be caused by density evolution of the galaxy population.
The density fit is used to correct the $1/V_{max}$ estimate. We also
demonstrate the high quality and completeness of our sample by
a variety of methods.

The luminosity function derived from this sample is compared to previously
published estimates, showing the prior estimates to have been strongly
affected by the Local Supercluster.  Our new luminosity function
leads to lower estimates of mid-infrared backgrounds and number counts.

\end{abstract}


\section{Introduction}
\label{sec:intro}

Much of the effort to study infrared-luminous galaxies has centered on
wavelengths greater than 50 $\mu$m.  Modeling work is focused on the
near-IR (e.g. \markcite{chok94}Chokshi et al. 1994) and far-IR (e.g.
\markcite{hac87}Hacking et al. 1987; \markcite{rrbinson96}Rowan-Robinson
et al. 1996) portions of the galaxian spectrum.
However, the mid-infrared is well-suited for studying starburst and
ultraluminous galaxies.  About 40\% of the luminosity from starburst
galaxies is radiated from 8-40 $\mu$m (\markcite{soi87}Soifer et al.
1987).  Extinction effects are small, and problems due to infrared
cirrus are minimized.  Most importantly for space astrophysics, for
a fixed telescope aperture, the spatial resolution is higher at shorter
wavelengths, and the confusion limit lies at higher redshifts.  Recent
work using the {\it Infrared Space Observatory (ISO)} (e.g.
\markcite{knapp96}Knapp et al.  1996; \markcite{boul96}Boulade et al.
1996; \markcite{rrobinson96}Rowan-Robinson et al. 1996)
shows the relative importance of the 7 $\mu$m and 15 $\mu$m bands
for galaxy studies.  The {\it Wide-Field Infrared Explorer (WIRE)},
a Small Explorer mission due to launch in late 1998
(\markcite{hac96}Hacking et al.\ 1996;
\markcite{schemb96}Schember et al. 1996), will conduct a very deep survey
at 24 $\mu$m to study
starburst galaxy evolution.  The {\it Space Infrared Telescope Facility
(SIRTF)} is also expected to conduct surveys in mid-infrared bands.
To interpret the results of these surveys now in
progress or soon to commence, it is necessary to better understand
the mid-infrared properties of galaxies in the local Universe.

The 25 $\mu$m luminosity function provides the basis
for predicting the
faint source counts in the mid-infrared.  The empirical model of
\markcite{hac91}Hacking \& Soifer (1991) uses an analytic fit to the
luminosity function derived by Soifer \& Neugebauer \markcite{soi91}
(1991).  This function was estimated from a complete subsample of the
Bright Galaxy Sample (\markcite{soi87}Soifer et al.\ 1987) containing
135 galaxies to a flux density limit of 1.26 Jy.  The availability of
many more redshifts of IRAS galaxies (principally from the 1.2 Jy Survey
(\markcite{str90}Strauss et al.\ 1990; \markcite{fis95} Fisher et al.\
1995)) enables a much larger sample to be studied, reducing uncertainties
at high and low luminosities.  

In this paper we present the selection of a large galaxy sample that is
flux-limited at 25 $\mu$m, and derive the local luminosity function
based on this sample.  The sample selection is described in the next
section.  In Section \ref{sec:lfresults} we describe the $1/V_{max}$ and
the maximum-likelihood estimators for deriving the local luminosity
function, and present the results. The completeness of the sample is
discussed in Section \ref{sec:compdisc}. In Section \ref{sec:correct} we
calculate the radial density distribution of the sample using a
maximum-likelihood method. The radial density fit is used to correct the
$1/V_{max}$ estimate of the local luminosity function, as well as the
redshift distribution with which the luminosity function can be
compared.  Section 6 includes discussions of the different luminosity
function estimators, a comparison of our newly derived luminosity
function with previous estimates, the implications for mid-infrared
backgrounds and number counts, and the effects of evolution on the
derivation of the luminosity function. The color properties of the
sample are treated in another paper (\markcite{fang98}Fang et al.\ 1998).

\section{Sample Selection}
\label{sec:sample}

Our sample is based on a selection from the IRAS Faint Source Survey
(FSS;\markcite{mos92} Moshir et al.\ 1992).  The principal data product
of the FSS is the Faint Source Catalog (FSC).  The FSC was produced
by coadding IRAS detector scans before extracting sources, and
is roughly one magnitude deeper than the Point Source Catalog.
Another FSS database is the Faint Source Reject File (FSCR), which contains
possible detections that were not included in the FSC for assorted
quality-control reasons.  This section treats the selection of sources
from the FSC and the FSCR in turn.

To minimize the effects of cirrus and contamination from stars, we chose a
sky coverage limit of $|b| \ge 30^\circ$.  In addition,
the same excluded areas used by Strauss et al.\ (1990) were excluded
from our sky coverage region.  At high Galactic latitudes, the
excluded regions consist almost wholly of gaps not covered by IRAS,
and the Large and Small Magellanic Clouds.  Our sample covers
47.7\% of the sky.

The 90\% completeness limit for the FSC lies at a 25 $\mu$m flux
density of 210 mJy for sky covered by 2 HCONs, and at 170 mJy for
coverage of 3 HCONs, for $|b| > 10^\circ$ (Moshir et al.\ 1992).  Our
preliminary searches of redshift databases indicated, however, that
for such faint flux density limits, many of the faintest objects
would have no redshift available.  A flux density limit of 250 mJy
was found to have a small redshift incompleteness (see below) and was
selected for our sample.  Furthermore, sources with moderate (SNR=3
to 5) or high-quality (SNR $> 5$) flux densities at 25 $\mu$m are
included in the sample.

A set of weak color criteria were chosen to further discriminate against
stars.  The great majority of galaxies have larger flux density at 60
$\mu$m than at 25 $\mu$m.  For these sources, no constraint was placed
on the 12 $\mu$m -- 25 $\mu$m color.  A small percentage of galaxies
will have a smaller flux density at 60 $\mu$m than at 25 $\mu$m.  For
these sources, we have limited the ratio of $F_{\nu}(25)/F_{\nu}(60)$
to lie between 1 and 1.6, and have further constrained
$F_{\nu}(12)/F_{\nu}(25)$ to be less than 1 to avoid admitting
large numbers of stars into the sample.  Color-color diagrams of
the sources in the sample that have been identified as galaxies with
redshifts (see below) are displayed in Figure \ref{fig:colorcolor}.
These diagrams show that only a handful of galaxies
satisfy $F_{\nu}(25)/F_{\nu}(60) \ge 1$, and that our color criteria
are fairly liberal for sources identified as galaxies.

To summarize, our sample selection criteria are:
\[ \begin{array}{l}
    F_{\nu}(25) \ge 250  {\rm mJy;} \\
    \mbox{moderate or good quality detection at}~25\mu{\rm m;} \\
    |b| \ge 30^\circ, \mbox{ and not in Strauss et al.\ excluded zone}; \\
    F_{\nu}(25)/F_{\nu}(60) < 1 \mbox{ and no constraint on } F_{\nu}(12)/F_{\nu}(25), \\
     {\rm or} ~ 1  < F_{\nu}(25)/F_{\nu}(60) < 1.6 ~{\rm and}
                               ~F_{\nu}(12)/F_{\nu}(25) < 1.
\end{array} \]

1619 sources in the FSC meet the above criteria.  To identify galaxies and
non-galaxies (stars, infrared cirrus, etc.), and to obtain redshifts for
identified galaxies, a positional match of these 1619 sources was made with
the 1.2 Jy Survey catalog, the November 1993 public version of J.P. Huchra's
ZCAT, the NED database, and the SIMBAD database.  A matching radius of 1
arcminute was used to match the FSC sources with these catalogs.  Additional
redshifts were culled from J.P. Huchra's private ZCAT catalog, and from
additional observations made at the Fred L. Whipple Observatory (FLWO) 1.5m.
This matching process resulted in 1438 galaxies with redshifts.  Of the
remainder, 166 sources were identified as Galactic objects or H{\sc II}
regions in nearby galaxies, leaving 15 sources identified as galaxies without
redshifts.

The selection of sources from the FSCR is more complicated due to the
low reliability of sources in that database.  Applying our selection
criteria to the Faint Source Reject File results in 851 sources.  These
were matched with the IRAS OPTical IDentification (OPTID) database at
IPAC.  The OPTID database is a special version of the FSC and FSCR with
optical identifications added from the Guide Star Catalog (GSC:
\markcite{lasker90}Lasker et al.\ 1990), the Tycho Input Catalog (TIC:
\markcite{egr92}Egret et al.\ 1992)), and the COSMOS/UKST Southern Sky
Object Catalogue (COSCAT: \markcite{yen92}Yentis et al.\ 1992).  Sources
matching with stars within $1\sigma$ error ellipses on the OPTID Plots
were excluded. We examined those sources which have no match with either
a star or a galaxy within $1\sigma$ error ellipses by studying the OPTID
Reports and by comparing with the optical Digitized Sky Survey images. 
We found only a few of these sources to be possible galaxies which,
along with the reject sources that match with galaxies within the
$1\sigma$ error ellipses, gives a small sample of 27 sources.  Among
these, there are 5 sources which are extremely blue, and have null SNRs
at 60 $\mu$m.  ADDSCANS (one-dimensional coadds of IRAS scans) of
these sources show no obvious detections at their locations at all four
IRAS bands.  No redshifts for these sources were found when matching
them with the redshift catalogs (see above), and we suspect that they
are not real galaxies.  The final yield from the Reject File is 22
sources.

Matching the 22 FSCR sources with the same redshift catalogs gives 20
sources with redshifts.  To verify that these matches are not random, a
sample of 851 sources randomly distributed in our sampling regions was
generated.  The number of matches with galaxies is only 2.  This
indicates that these 22 reject sources are not likely to be just random
matches with galaxies.  Furthermore, it shows that the number of
coincidental matches in our total sample should also be very small.

The breakdown of sources for redshifts for our entire sample (FSC plus
FSCR sources) is listed in Table \ref{tab:redsrc}.  The previously
unpublished redshifts obtained at FLWO are tabulated in Table
\ref{tab:redshifts}. The total number of galaxies with redshifts is
1458. There are 17 galaxies remaining without redshifts.

The 1.2 Jy Survey catalog also provides the ADDSCAN flux densities for the
sources extended at 60 $\mu$m.  For those sources which have no ADDSCAN
flux densities in the 1.2 Jy Survey but are extended at 25 $\mu$m, we have
obtained new ADDSCAN flux density measurements and substituted these for the
FSC values.

Figure \ref{fig:ncounts} shows the $\log(N)$ versus $\log(F_{\nu}(25))$ plot
for the sample of galaxies with redshifts.  The points follow a slope of -1.5
all the way to our flux limit, demonstrating the high completeness of our
sample as a function of flux.  The apparent excess at high flux values is most
likely due to the Local Supercluster.  Further checks of completeness are
presented in Section \ref{sec:compdisc}, including an investigation of
possible systematic effects of flux errors on derived luminosity functions.

For each galaxy, secondary distances are used when available. 
Otherwise, distances have been assigned using a linear Virgocentric
inflow model (\markcite{aar82}Aaronson et al.\ 1982) and a Hubble
constant of 75 km~s$^{-1}$ Mpc$^{-1}$ (see Appendix \ref{sec:h0disc} for
the case of a different Hubble constant).  The same criteria as Soifer
et al. \markcite{soi87} (1987) have been used to identify Virgo cluster
galaxies (with a distance of 17.6 Mpc).  The monochromatic luminosity of
each source is computed from the distance, using a $k$-correction that
assumes a power-law slope of the SED between 12 $\mu$m  and 25 $\mu$m. 
The monochromatic luminosities are expressed as $\nu L_\nu$ and have
units of solar luminosities.

\section{Luminosity function results}
\label{sec:lfresults}

\subsection{1/$V_{max}$ and parametric maximum likelihood}
\label{sec:parametric}

The classic method of estimating luminosity functions is via
the $1/V_{max}$ estimator (Schmidt 1968).  This technique is
non-parametric, so no analytic form of the luminosity function
need be assumed, but the data must be binned.  The infrared
luminosity functions presented by Soifer et al.\ (1987) and
Soifer \& Neugebauer (1991) were derived via this method.
The space density $\rho_L$ and its error $\sigma_{\rho}$ are
computed from the following quantities:
\begin{equation}
\rho_L = \left ( {{4\pi} \over \Omega} \right )
        \left ( {\sum {1 \over {V_{max}}}} \right ) ,
\end{equation}
\begin{equation}
\sigma_{\rho} = \left ( {{4\pi} \over \Omega} \right )
        {\left ( {\sum {1 \over {V_{max}^2}}} \right ) }^2 ,
\end{equation}
where $\Omega$ is the solid angle of the survey, $V_{max}$ is the
maximum volume to which the object could have been detected, and the
sum is over all galaxies in each luminosity bin.  The assumption
that galaxies are
distributed uniformly in space may be tested by checking that
$V/V_{max}$ is 0.5 in each luminosity bin.

The luminosity function may also be derived via maximum likelihood
techniques that are independent of density variations (e.g.\markcite{san79}
Sandage, Tammann, \& Yahil 1979). The parametric method has the
advantage that no binning of the data is required, but an analytic form
of the luminosity function must be assumed. For all of the parametric
maximum likelihood fits in this paper, we have assumed a cumulative luminosity
function of the form used by Yahil et al.\ \markcite{yah91} (1991) in
the analysis of the 1.2 Jy Survey:
\begin{equation}
\Psi(L)=C{\left({{L}\over{L_*}}\right)}^{-\alpha}{\left( 1+{{L}\over{L_*}}
                                                 \right)}^{-\beta}
\end{equation}
with its corresponding differential luminosity function
\begin{equation}
\Phi(L)={\left({\alpha \over L} + {\beta \over 
                        {L_* + L}}\right)}\Psi(L).
\end{equation}

A flux-limited sample contains only a small number of very sub-luminous
galaxies, making it difficult to determine the luminosity function at
these luminosities.  We have followed Yahil et al.\
\markcite{yah91}(1991) 
in imposing a lower limit on luminosity $L_s = 4 \pi r_s^2 \nu
f_m$, where $r_s$ is the distance corresponding to a velocity of 500 km
s$^{-1}$, and $f_m$ is the survey flux limit.  For $f_m$ = 250 mJy,
$L_s$ is $4.0\times 10^7 L_\odot$. Then the minimum
detectable luminosity at distance $r$ is $L_{min}(r) = {\rm Max} (L_s,
4\pi r^2 \nu f_m)$.  The probability of detecting a galaxy of
luminosity $L_i$ at distance $r_i$ is then
\begin{equation}
\label{eqn:condprob}
f(L_i | r_i) = \left \{ \begin{array}{ll}
        {{\Phi(L_i)}/{\Psi(L_{min})}} & \mbox{if $L_i \ge L_{min}$} \\
        0 & \mbox{otherwise}
         \end{array}
         \right .
\end{equation}
The likelihood function is the product of all of these probabilities:
\begin{equation}
\Lambda = - 2 \sum_i \ln f(L_i | r_i).
\end{equation}
This (reciprocal) likelihood function is minimized to determine 
$\alpha$, $\beta$, and 
$L_*$.  The variance of these parameters is estimated by making
use of the asymptotic
normality of the maximum likelihood estimators for a large sample 
(\markcite{kendall87}Kendall, Stuart \& Ord 1987).

The likelihood function contains ratios of the differential and cumulative
luminosity functions, so another calculation is required to find 
the normalization $C$.
One method is to use
the $n_1$ estimator of Davis \& Huchra \markcite{dav82}(1982) after
the shape parameters have been determined.  The quantity $n_1$ is the
number density of galaxies with $L \ge L_s$.  (See Equations 11 and
13 of \markcite{yah91}Yahil et al.\ (1991) for the formulas appropriate
to our choice of the parametric luminosity function.) An upper limit
to the redshift must be chosen to fix the sample volume used in this
density estimator.  Based on a calculation of $n_1$ in radial shells, 
we have
chosen an upper limit on redshift of 20,000 km~s$^{-1}$ ($z \sim 0.067$)
for the normalization of the maximum likelihood fit.  (However, this
upper redshift limit is {\it not} applied in determination of the
shape of the luminosity function.)  Effects of  incompleteness or
evolution may become significant beyond this redshift.
A maximum-likelihood fit to the density
variations of the sample as a function of redshift (see Section
\ref{sec:correct}) further indicates that the density is well-behaved
to this limit.  A second means of determining the normalization is to
require that the number of sources predicted by the luminosity function
over some redshift range equals the number in the sample.  Computing
the normalization from source counts in this way from 500 km s$^{-1}$
to 20,000 km~s$^{-1}$ gives essentially the same normalization as
the $n_1$ estimator.

The derived parameters for the maximum likelihood
fit are listed in Table \ref{tab:fitpars}.
Luminosity functions derived by the $1/V_{max}$ method and the
parametric maximum likelihood method are shown in Figure
\ref{fig:newvis}.  We show the luminosity
functions in the form of the visibility $\Theta = \Phi L^{2.5}$.
The visibility is proportional to the number of galaxies visible
per log luminosity bin,
and it has the advantage of showing the relative numbers of galaxies
in a flux-limited sample.  The open triangles are from the $1/V_{max}$
estimate of the entire sample, and the solid curve is the parametric
maximum-likelihood fit to galaxies with redshifts greater than 
500 km~s$^{-1}$, with normalization from a volume-limited sample
as described above.    The solid curve is truncated at
$L_s = 4.0\times 10^7 L_\odot$.  The $1/V_{max}$ estimate (made in bins
of $\log(\nu L_\nu) = 0.4$) was converted to visibility by
multiplying $\rho_L$ by  $L^{1.5}{2.5 \over {\ln 10}}$.

A plot of $V/V_{max}$ as a function of luminosity for our sample is
included in Figure \ref{fig:vratio1}.  The variations in this
statistic for the full sample and northern sample at luminosities
less than $10^{10} L_\odot$ are most likely due to the Local
Supercluster.  The southern sample is much more uniform at these
luminosities, supporting this interpretation.

The general agreement between the $1/V_{max}$ luminosity function and
the shape of the parametric fit in Figure \ref{fig:newvis}, together
with the uniformity of the $V/V_{max}$ statistic, confirm that the
parametric fit is a good description of the shape of the local
luminosity function.    The $1/V_{max}$ point at $\log \nu L_\nu = 8.6$
corresponds to the luminosity at which the flux limit samples the
Virgo cluster, and hence lies well above the solid curve.  

The number of galaxies without redshifts (17) is slightly more than one
percent of our sample.  One method for estimating the effects of this known
incompleteness on the luminosity function is to assign to each of these
galaxies the median redshift of their log flux bin, and recalculate the
luminosity function.  When the median redshifts are computed for the whole
sample, all the assigned redshifts are around 7,000 km sec$^{-1}$.  When the
median redshifts are calculated from those galaxies whose redshifts were not
obtained from public sources (a more realistic estimate), the assigned
redshifts range from 12,000 to 18,000 km sec$^{-1}$.  Since all these galaxies
have a 25 $\mu$m flux density within a factor 1.5 of the flux limit, their
luminosities all fall near $L_*$.  When the luminosity function is
recalculated and plotted as a visibility, we have found that the asymptotes are
unchanged, and the peak visibility is increased by a few percent.

\subsection{Nonparametric maximum likelihood}
\label{sec:lfnonpar}

A non-parametric maximum-likelihood method (a.k.a ``Stepwise Maximum
Likelihood'') has been used to calculate the galaxy luminosity function in
various surveys (\markcite{lyn71}Lynden-Bell 1971;
\markcite{cho86}Choloniewski 1986; \markcite{efs88}Efstathiou et al.\ 1988;
\markcite{sau90}Saunders et al.\ 1990; \markcite{marzke94}Marzke et al. 1994).
This method does not assume a specific functional form for the luminosity
function, but rather describes it by a non-negative step function
(i.e. requires binning) which makes the observed galaxy luminosity
distribution as likely as possible.  We have followed the prescriptions
set out by \markcite{efs88}Efstathiou et al.\ (1988) and 
\markcite{sau90}Saunders et al.\ (1990) for calculating the non-parametric
luminosity function.

As in the parametric maximum-likelihood method, the non-parametric approach
also eliminates the effect of any density inhomogeneities, but loses the
overall normalization by assuming that the galaxy spatial and luminosity
distributions are independent.  The
normalization of the non-parametric function has been chosen to give the
same source-counts as the parametric function over the same bins.

Figure \ref{fig:lf_swml} shows the visibility function obtained by using
the non-parametric maximum-likelihood method (circles), which agrees within
the error bars with our parametric results (solid line, from Figure
\ref{fig:newvis}).    
The non-parametric method is commonly used to assess whether the parametric
functional form is acceptable.  Figure \ref{fig:lf_swml}
shows that the parametric form of Figure \ref{fig:newvis} gives a reasonable
description of the density-independent 25 $\mu$m luminosity function.
We have found that the non-parametric function is sensitive to the bin
width, however.  Figure \ref{fig:lf_swml} shows the case for bins in
$\log L$ = 0.2, but for a bin width of 0.4, the non-parametric results is
systematically higher than the parametric curve for high luminosities.
The coarser binning probably results in a loss of information and a less
reliable luminosity function.  The choice of bin size is a compromise
between this effect and tolerating the statistical uncertainty from
having too few sources in each bin.

\section{Completeness Checks}
\label{sec:compdisc}

Systematic variations of noise across the sky may change the
detection quality and cause incompleteness problems for the sample.
In the 25 $\mu$m passband of IRAS, noise is dominated by photon shot noise
from the zodiacal background, and hence is a strong function of the
ecliptic latitude.  We have examined the noise variation in the
sample and in the corresponding FSS plates and found two regions that
contain higher noise at $25\mu$m.  They are essentially the two
quadrants defined by $(0^\circ < l < 180^\circ, b < -30^\circ)$ and
$(180^\circ < l < 360^\circ, b > 30^\circ)$, which contain the ecliptic 
plane.  Most of the moderate-quality detection sources (generally,
those with SNRs between 3 and 5) in our
sample are in these two regions, indicating the systematic effect.

To estimate this systematic effect on our sample completeness, we
compared luminosity functions estimated for two subsamples of our
250-mJy-limited full sample.  The first subsample is defined by raising
the flux-density limit to 400 mJy in the two quadrants of the sky
$(0^\circ < l < 180^\circ, b < -30^\circ)$ and $(180^\circ < l <
360^\circ, b > 30^\circ)$ where the noise is systematically higher, and
keeping the 250 mJy limit in the other two quadrants.  Most of the
moderate quality detection sources are left out of this subsample, so we
call these 1050 sources the high-quality subsample.  The second
subsample consists of those 408 sources in the full sample that are left out
of the high-quality subsample, and is therefore a low-quality subsample.
Figure \ref{fig:comphighlowQ} compares the luminosity functions of the
high-quality and the low-quality subsamples, using the $1/V_{max}$
estimator since it accomodates the different flux limits more easily
than either of the maximum-likelihood estimators.  There is no
systematic deficiency of galaxies in the low-quality subsample.  The
luminosity functions agree within the $1\sigma$ error bars, with the
exception of the point at $\log \nu L_\nu = 8.6$.  The low-quality
subsample includes galaxies in the Virgo cluster with 25 $\mu$m
flux densities between 250 and 400 mJy, most of which fall into this
luminosity bin.

In Figure \ref{fig:comphighQwhole} we compare the $1/V_{max}$ luminosity
functions of the high-quality subsample and the full 250 mJy-limited sample.
They agree very well, indicating that the full sample is complete at 250 mJy
flux-density across the entire sampling sky.  Therefore we are confident
that the full 250 mJy-limited sample is well-defined.

As another check of the completeness of our full sample, we have computed
the cumulative probability of finding a galaxy of luminosity $L$
at distance $r$, following the method of Yahil et al.\
\markcite{yah91} (1991).  This probability is the integral of
Equation \ref{eqn:condprob} and is $F(L|r)=\Psi(L)/\Psi(L_{min}(r))$.
This probability should be uniformly distributed on the interval
[0,1] independent of distance.  If the sample were incomplete at
the faint end, we would expect a decrease in the relative number
of galaxies as $F\rightarrow 1$.

We have computed $F(L_i | r_i )$ for each galaxy in the sample, then
computed the histogram of this statistic in different distance ranges,
normalized to 1 in each range.  The result is shown in Figure
\ref{fig:compcheck}.
There is no systematic decrease in the relative galaxy counts as
$F\rightarrow 1$ (albeit the statistics are poor beyond 400 Mpc).
This result provides further confirmation that our 250 mJy-limited sample
is complete to these distances.  Since this method relies upon ratios
of the luminosity function, it should be insensitive to inhomogeneities
in the galaxy distribution.

\section{Density Variations and the Corrections for $V/V_{max}$}
\label{sec:correct}

Just as the galaxy luminosity function independent of the
inhomogeneities of the density field can be obtained, the spatial
density variations independent of the galaxy luminosity function can
also be calculated using the maximum-likelihood method
(\markcite{sau90}Saunders et al.\ 1990).  Here we only investigate the
radial density distribution.  The probability for detecting a source of
luminosity $L_{i}$ at radial distance $r_{i}$ is
\begin{equation}
p_{i}=\frac{\rho(r_{i})dV_{i}}{\int_{0}^{r_{max,i}}\rho(r)(dV/dr)dr},
\end{equation}
where $\rho(r_{i})$ is the density at $r_{i}$, and $r_{max,i}$ is the
maximum distance for the source to be included in the sample.  The
likelihood function for all the detected sources would be $\cal L = \prod
\it p_{i}$, \rm given the density distribution $\rho(r)$ we are trying to
calculate.  In the non-parametric approach, this distribution can be found
by maximizing $ln \cal L$ with respect to the binned density field
$\rho(r_{j})$, which gives
\begin{equation}
\label{eqn:bindensity}
\rho(r_{j})=N_{j}(\sum_{i}\frac{f_{ij}V_{j}}{V_{eff,i}})^{-1},
\end{equation}
where $V_{eff,i} = \sum_{k=0}^{r_{max,i}} \rho(r_{k})V_{k}$,
$N_{j}$ is the number of sources in the $j$-th distance bin with volume
$V_{j}$, $f_{ij}$ is the fraction
of the bin contained in $[0, r_{max,i}]$ where the $i$-th source is observable,
and the sum is over all sources.

Figure \ref{fig:den_swml} illustrates the non-parametric maximum-likelihood
estimate of the radial density variation (solid line) as compared with the
flux-limited observed distribution (dashed line).  Here the radial distance
bins are in logarithmic scale, and the density variations are normalized to
the average density over all bins of interest.
The method correctly recovers the density enhancement at $\sim
17$ Mpc caused by Virgo.

We do not find a systematic increase in the radial density at large
distances.  If our sample is assumed to be complete, this result means
that there is no indication of density evolution in this sample.
We return to this point in section \ref{sec:evolution}.

The benefits of calculating the radial density variations in our sample
are two-fold.  First, it may be used to correct the $V/V_{max}$
statistic and the traditional $1/V_{max}$ estimate of the
density-dependent luminosity function (\markcite{sch68}Schmidt 1968;
\markcite{fel77} Felton 1977).  For the $V/V_{max}$ test, since
\begin{equation}
\frac{\int_{0}^{r_{max}}\frac{\int_{0}^{r}\rho(r')D^{2}dr'}
{\int_{0}^{r_{max}}\rho(r')D^{2}dr'}\rho(r)D^{2}dr}
{\int_{0}^{r_{max}}\rho(r)D^{2}dr} =
\frac{\int_{0}^{r_{max}}(\int_{0}^{r}\rho(r')D^{2}dr')d(\int_{0}^{r}\rho(r)
D^{2}dr)}{(\int_{0}^{r_{max}}\rho(r)D^{2}dr)^{2}} = \frac{1}{2},
\end{equation} where $\rho(r)$ is the radial density field, $r_{max}$ is
the maximum comoving distance for a source to be included in a sample,
and $D$ is the effective distance (\markcite{lon78}Longair 1978), we can
modify the $V/V_{max}$ test so that \begin{equation}
\label{eqn:modvovmax}
\frac{V}{V_{max}}=\frac{\int_{0}^{r}\rho(r)D^{2}dr}
{\int_{0}^{r_{max}}\rho(r)D^{2}dr}, \end{equation} which gives
$\overline{V/V_{max}}=1/2$ for an inhomogeneous radial distribution. 
This also suggests that we use the $V_{eff,i}$ in equation
\ref{eqn:bindensity} instead of $V_{max}$ in the $1/V_{max}$ estimator
to calculate a density-corrected luminosity function.  The $\rho(r)$
calculated from the maximum-likelihood method can provide these
corrections.  The results are shown in Figure \ref{fig:vmax2} for the
new $V/V_{max}$ values and in Figure \ref{fig:lfvmax} for the
$1/V_{max}$ luminosity function.  Compared with the results without the
inhomogeneity correction in Figure \ref{fig:vratio1}, the $V/V_{max}$ values
are closer to 0.5 for those bins most strongly affected by
the Local Supercluster.  (However, the
corrected $V/V_{max}$ values are not suitable as a test of the sample
completeness, since the density correction will mask any incompleteness
in the sample.) The new $1/V_{max}$ luminosity function is smoother and
closer to the ones estimated by the maximum-likelihood method.  As
mentioned in Saunders et al. (1990), the binning of the distance in
Figure \ref{fig:den_swml} may account for the residual effects of the
density inhomogeneities in these Figures.

Second, the ``true'' density variations in Figure \ref{fig:den_swml}
also correct the observed redshift distribution. The
density-variation-corrected redshift distribution should be predicted by
a density-independent luminosity function estimated in the same sample. 
This provides an independent test to verify the luminosity function. 
Figure \ref{fig:srccnt} shows our results.  Here the dashed histogram is
the observed redshift distribution. (Here we show the source-counts in
distance bins instead of redshift bins since some galaxies have
secondary distances.) It is divided by the density of the corresponding
distance bins in Figure \ref{fig:den_swml} which gives the solid-line
histogram.  The source-counts affected by large-scale structures, such
as the ones near Virgo, are corrected by this procedure. The dotted line
shows the prediction by the parametric density-independent luminosity
function of Figure \ref{fig:newvis} and Table \ref{tab:fitpars}.  This
prediction fits the solid-line histogram well, demonstrating the
reliability of our luminosity function.  Both the solid-line histogram
and the dotted line are normalized so that they contain the same total
number of sources as observed within 20,000 km~s$^{-1}$.

\section{Discussion}
\label{sec:disc}

\subsection{Limitations of the methods}
\label{sec:limits}

The preceding sections have presented the 25 $\mu$m luminosity function
calculated by a number of methods.  In this section, the limitations of
the various methods are revisited in greater detail.

The $1/V_{max}$ estimate of the luminosity function varies the most from
the other estimators, particularly at luminosities less than $L_*$, where
our sample is most strongly influenced by the Local Supercluster.  As
noted above, this estimator assumes that the 3-dimensional galaxy
distribution is homogeneous.  The error bars in all of our $1/V_{max}$
estimates similarly depend on the assumption of uniformity, and do not
reflect the additional uncertainties that arise from an inhomogeneous
distribution.  We consider the $1/V_{max}$ estimates to be the least
reliable descriptions of the 25 $\mu$m luminosity function out of those
we have presented.

The parametric and non-parametric maximum-likelihood estimates should
be independent of structures in the galaxy distribution.  However,
these estimators assume that the luminosity function is independent
of density.  This assumption is difficult to check.  In the case of a
large 60 $\mu$m-selected sample, \markcite{yah91} Yahil et al. (1991)
demonstrated that their similarly-derived luminosity function
provided a good description of the luminosity distribution of
galaxies even in regions of high density.

The non-parametric maximum likelihood method applied to our sample is
sensitive to the choice of binning, as described in section 
\ref{sec:lfnonpar}. We therefore consider the parametric
maximum-likelihood description of the 25 $\mu$m luminosity function with
parameters in Table \ref{tab:fitpars} to be the best out of those we
present in this paper.

\subsection{Comparison with other luminosity functions}
\label{sec:compare}

As mentioned in the Introduction,
\markcite{soi91}Soifer \& Neugebauer (1991) derived the local luminosity
function at 25 $\mu$m from a complete subsample of the 60-$\mu$m-selected
Bright Galaxy
Sample (\markcite{soi87}Soifer et al.\ 1987).  This subsample contained
135 galaxies to a flux density limit of 1.26 Jy, and the luminosity function
was determined using the classical $1/V_{max}$ estimator. Their luminosity
function is plotted as a space density in Figure \ref{fig:spacedens}, 
together with our parametric and $1/V_{max}$ estimates. 

Our flux density limit of 250 mJy samples the Virgo cluster at $\log \nu
L_\nu = 8.6$ (section \ref{sec:parametric}), and our $1/V_{max}$ estimate
is much higher than the parametric curve in that bin. The Soifer \&
Neugebauer points in Figure \ref{fig:newvis} in $\log \nu L_\nu = 9.4$
bin is also much higher than the curve, as expected since their flux
density limit is five times higher than ours.  We conclude that the
Soifer \& Neugebauer luminosity function is strongly affected by the
Local Supercluster, particularly because the Bright Galaxy Sample  was
selected to be visible from Palomar Observatory (Soifer et al.\ 1984)
and does not include the south Galactic cap.

\subsection{Implications}
\label{sec:implications}

The improved luminosity function that is the main result of this paper
has implications the luminosity density of the local Universe, infrared
number counts predictions at 25 $\mu$m, and the infrared background
expected at this wavelength.

\markcite{soi91}Soifer \& Neugebauer (1991) calculated the local luminosity
density in the four IRAS passbands based on their derived luminosity
functions.  At 25 $\mu$m, they found that the log of the luminosity density in
units of $L_\odot$ Mpc$^{-3}$ is 7.2.  Integrating our parametric function
gives a log luminosity density of 6.9, which is a factor of 2 lower in linear
terms than the Soifer \& Neugebauer result. The differences in the sub-$L_*$
portion of the luminosity functions account for the different integrated
luminosity densities.

\markcite{hac91} Hacking \& Soifer (1991) presented number count
distributions at 25 $\mu$m, using a fit to the \markcite{soi91} Soifer
\& Neugebauer (1991) luminosity function.  (Those authors fit a
hyperbola to the  visibility function, which has the power-law
asymptotes at high and low luminosities like our parametric form, except
that it forces the absolute values of the slopes of the asymptotes to be
identical.) As they noted, the Soifer \& Neugebauer functions at other
wavelengths are 20\%-30\% higher than those derived from other samples
(\markcite{sau90}Saunders et al. 1990).  Our improved 25 $\mu$m function
is similarly about 15\% lower in normalization than the Soifer \&
Neugebauer estimate.  Hence, models based on this luminosity function
and its normalization will produce number counts about 15\% lower at the
bright end than those in Figure 1(c) of Hacking \& Soifer (1991).  Such
a change would bring the number counts in better agreement with the IRAS
number counts presented in their same figure.

Finally, we have substituted our luminosity function into the
backgrounds model of Hacking \& Soifer (1991).  The modified model
yields estimates of the infrared background at 25 $\mu$m that are
factors of 2.5 to 2.7 times smaller than the original Hacking \& Soifer
estimates.  Our results indicate that the detection of an IR
background at this wavelength is likely to be even more difficult
than previously predicted.

\subsection{Evolution}
\label{sec:evolution}

To this point, we have neglected any effects of galaxy evolution on our
derivation of the luminosity function.  Since the maximum-likelihood
estimators are independent of density variations, the shape of the
luminosity function will be unchanged by density evolution of the
galaxy population.  Our normalization considers galaxies out to 20,000 
km~s$^{-1}$, a redshift where evolution effects should be small.

More importantly, our maximum likelihood fit to the radial density
(section \ref{sec:correct} and Figure \ref{fig:den_swml}) does not show
a systematic increase with redshift and hence no indication of density
evolution.  This result is in direct contrast to that of 
\markcite{sau90}Saunders et al.\ (1990).  From their 60-$\mu$m-selected
sample, those authors found that density evolution of $(1+z)^{6.7}$
matched their radial density fit.  However, we note that the existence
of evolution from IRAS samples is not yet settled --- for the deepest
IRAS sample, \markcite{ashby96}
Ashby et al.\ (1996) did not find the high-redshift tail expected from
strong evolution.

Nevertheless, we have computed the shape of the parametric luminosity function
for the case of exponential luminosity evolution corresponding to $(1+z)^3$ at
low redshifts (cf. Saunders et al.\ (1990)), by correcting the luminosity of
each galaxy to the present epoch.  The values of the shape parameters of this
function are $\alpha = 0.437 \pm 0.032$, $\beta = 1.769 \pm 0.067$, and $L_* =
3.98 \pm 0.34 \times 10^9 L_\circ$.  The only change from the function
described in Table \ref{tab:fitpars} is that the slope of the luminosity
function at the high-luminosity end is slightly lowered.

\section{Summary and conclusions}
\label{sec:conclusions}

The following are the results of this paper:

1. We have selected a sample of 1458 galaxies with redshifts 
from the IRAS Faint Source
Survey with a flux density limit of 250 mJy at 25 $\mu$m.  An additional
17 galaxies do not have redshifts available.

2. The local luminosity function is derived using the $1/V_{max}$ estimator
and both parametric and non-parametric maximum likelihood methods.  
The $1/V_{max}$ estimate
is significantly affected by the Local Supercluster.  The maximum likelihood
methods are independent of density variations, and we consider the parametric
fit with parameters in Table \ref{tab:fitpars} to be the best estimate of the
local luminosity function at 25 $\mu$m.

3. A maximum likelihood fit to the radial density in this sample is used
to correct the $1/V_{max}$ estimate.  The fit shows no sign of a 
systematic increase with redshift of the density, as would result from
density evolution of the galaxy population.

4. The $1/V_{max}$ luminosity function derived from a smaller sample by 
Soifer \& Neugebauer (1991) is significantly contaminated by the Local
Supercluster.  Predictions of number counts and local luminosity density
based on that function are 15-20\% higher than those indicated by our
improved luminosity function.  The new function also leads to lower
predictions of the mid-infrared background due to galaxies.

\acknowledgments 

We thank Susan Tokarz, Perry Berlind, and Jim Peters for observations 
made at F.L.\ Whipple Observatory.
We also thank Iffat Khan and Joe Mazzarella for assistance with database
searches, and Carol Lonsdale and Tom Soifer for helpful discussions.
We are grateful to Cong Xu for a critical reading of this manuscript.
We thank the anonymous referee and our Scientific Editor, Steven Willner,
for numerous comments that have improved this paper.
The SIMBAD database is maintained by CDS in Strasbourg, France.  The
NED database is supported at IPAC by NASA.

\appendix

\section{The Distance to Virgo and a Different Hubble Constant}
\label{sec:h0disc}

We have assumed a distance 17.6 Mpc to Virgo in the linear
Virgocentric inflow model (\markcite{aar82}Aaronson et al. 1982) to
calculate the distance to each galaxy which does not have a secondary
distance.  To be consistent with our assumption of 75 km~s$^{-1}$
Mpc$^{-1}$ for the Hubble constant, we have used 1019 km~s$^{-1}$ as
the observed velocity of Virgo corrected to the centroid of the Local
Group, and implicitly assumed an infall velocity of about 300
km~s$^{-1}$ for the Local Group toward Virgo.  The resulting Hubble
flow velocity of Virgo, $\sim$1320 km~s$^{-1}$, is consistent with
many other estimates (e.g.  \markcite{fab89}Faber et al. 1989;
\markcite{tonry90}Tonry et al.  1990; \markcite{mould95}Mould et al.
1995; \markcite{huchra96}Huchra 1996).  A different Hubble constant
would change the distance estimate to Virgo if the Hubble flow
velocity of Virgo remains unchanged.

We have calculated the $1/V_{max}$ luminosity function assuming no
Virgocentric inflows, in which we used the redshift as the direct
distance measure unless a secondary distance exists for a galaxy.
Figure \ref{fig:flowtst} compares the $1/V_{max}$ luminosity
functions for the 250 mJy-limited sample with and without Virgocentric
inflow.  A Hubble constant of 75 km~s$^{-1}$ Mpc$^{-1}$ is used in both
cases.  The two estimates agree very well.  This indicates
that the inflow model we use has a small impact on the estimate of
the distance of the galaxies in the sample, and that most of these
galaxies are not strongly affected by the gravity from Virgo.  Note
that a luminosity function simply scales with the Hubble constant
(the luminosity changes as $H_{0}^{-2}$)
if all distance estimates in a sample are from redshifts.

The number of galaxies with secondary distances in our sample is small
(only 22).  We have done experiments to verify that these galaxies
do not contribute statistically significantly to the calculated
luminosity function.  We therefore conclude that our results can be
simply scaled if a different Hubble constant is used.

\clearpage



\clearpage

\figcaption[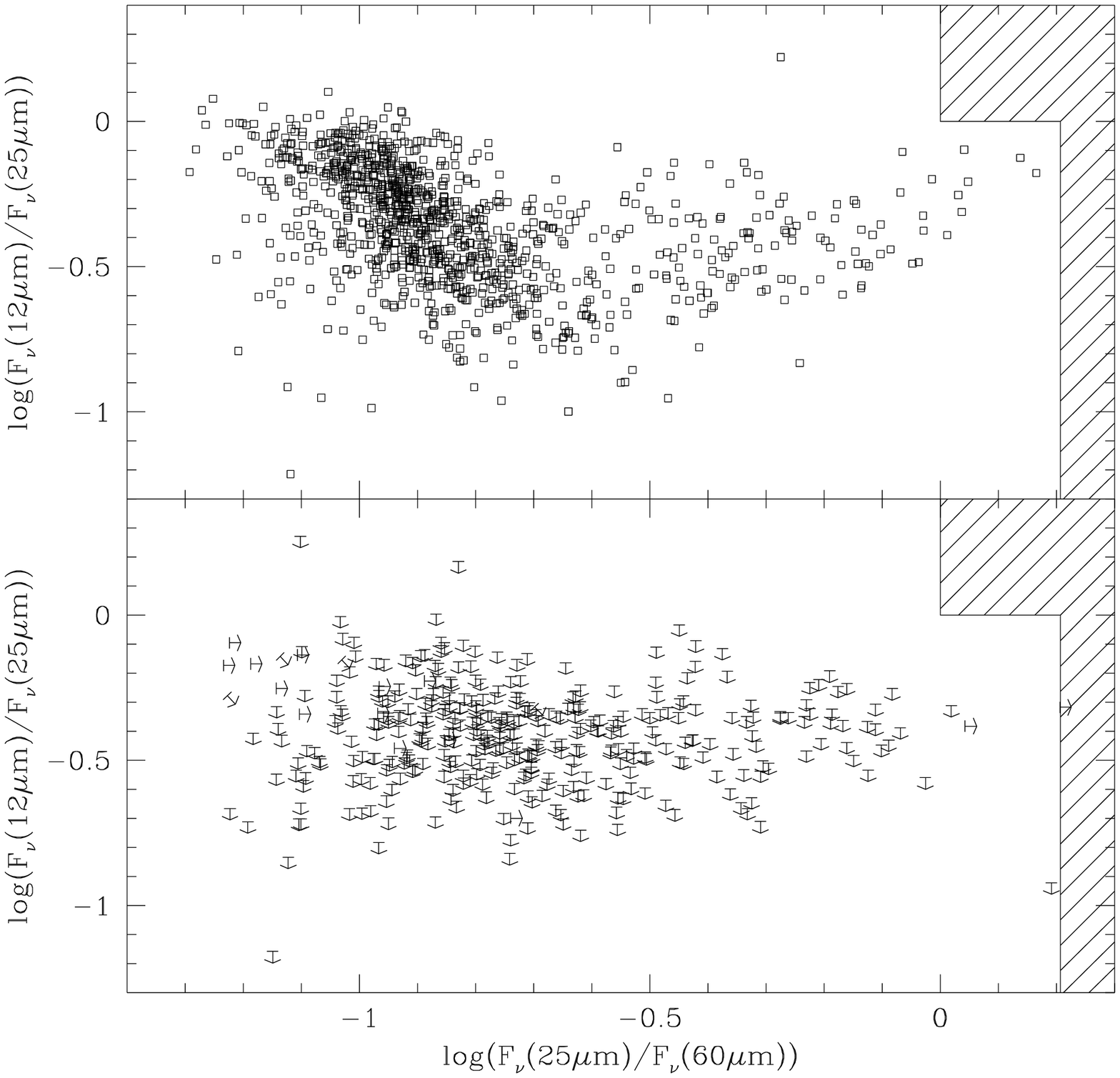]{ 
Color-color plots for the galaxies with redshifts in the sample.  The
upper plot shows galaxies with good or moderate detections in each of
the 12, 25 and 60 $\mu$m bands.  The lower plot show galaxies with
upper limits in either or both the 12 or 60 $\mu$m bands.  The
hatched region shows 
the region of the color-color
diagram that is excluded from the sample selection.
\label{fig:colorcolor}}

\figcaption[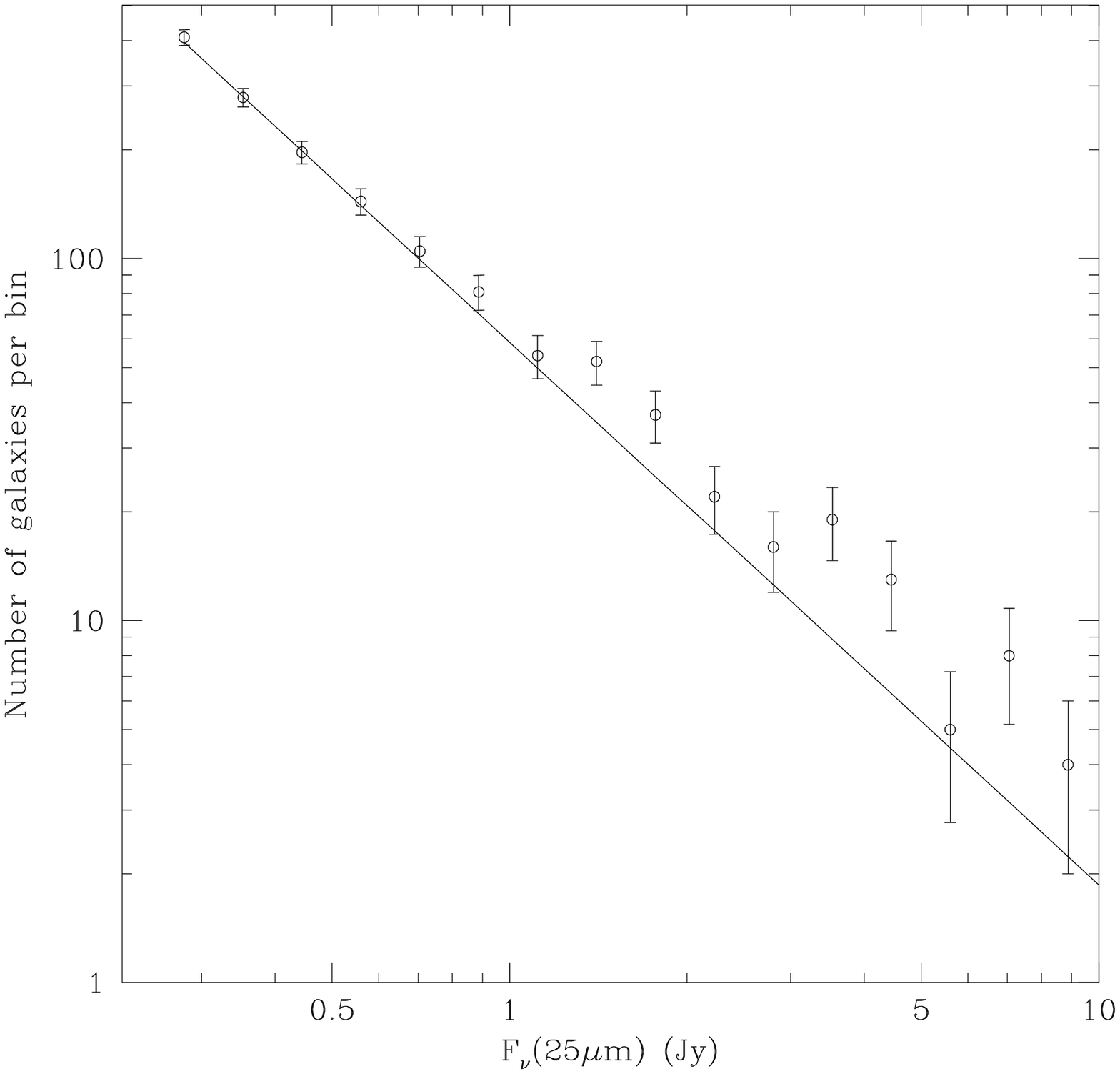]{
Differential number counts versus flux for the galaxies with 
redshifts in the sample. The line has a slope of -1.5.
\label{fig:ncounts}} 

\figcaption[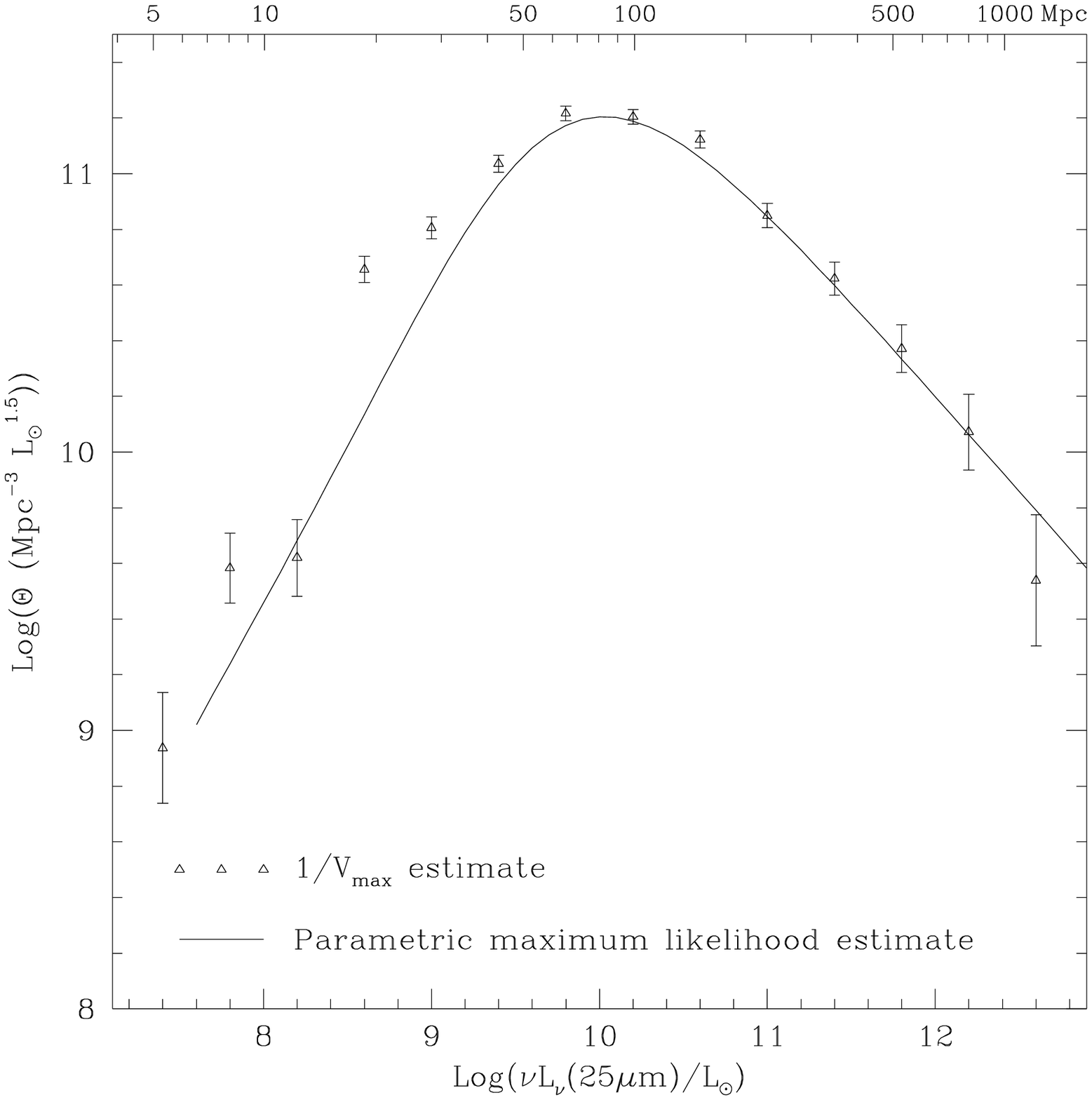]{
Visibility plots for the 250 mJy sample.  The open triangles represent
the results of the $1/V_{max}$ estimate of the luminosity function fit
to the entire sample.  The solid line is the parametric maximum
likelihood estimate for galaxies with redshifts greater than $500$
km~s$^{-1}$.  The parameters of the luminosity
function are listed in Table \ref{tab:fitpars}.  
The curve is truncated at $L_s =
4\times 10^7 L_\odot$. The distance scale at
top is computed from the 250 mJy flux density limit and the luminosity
scale at the bottom.
\label{fig:newvis}}

\figcaption[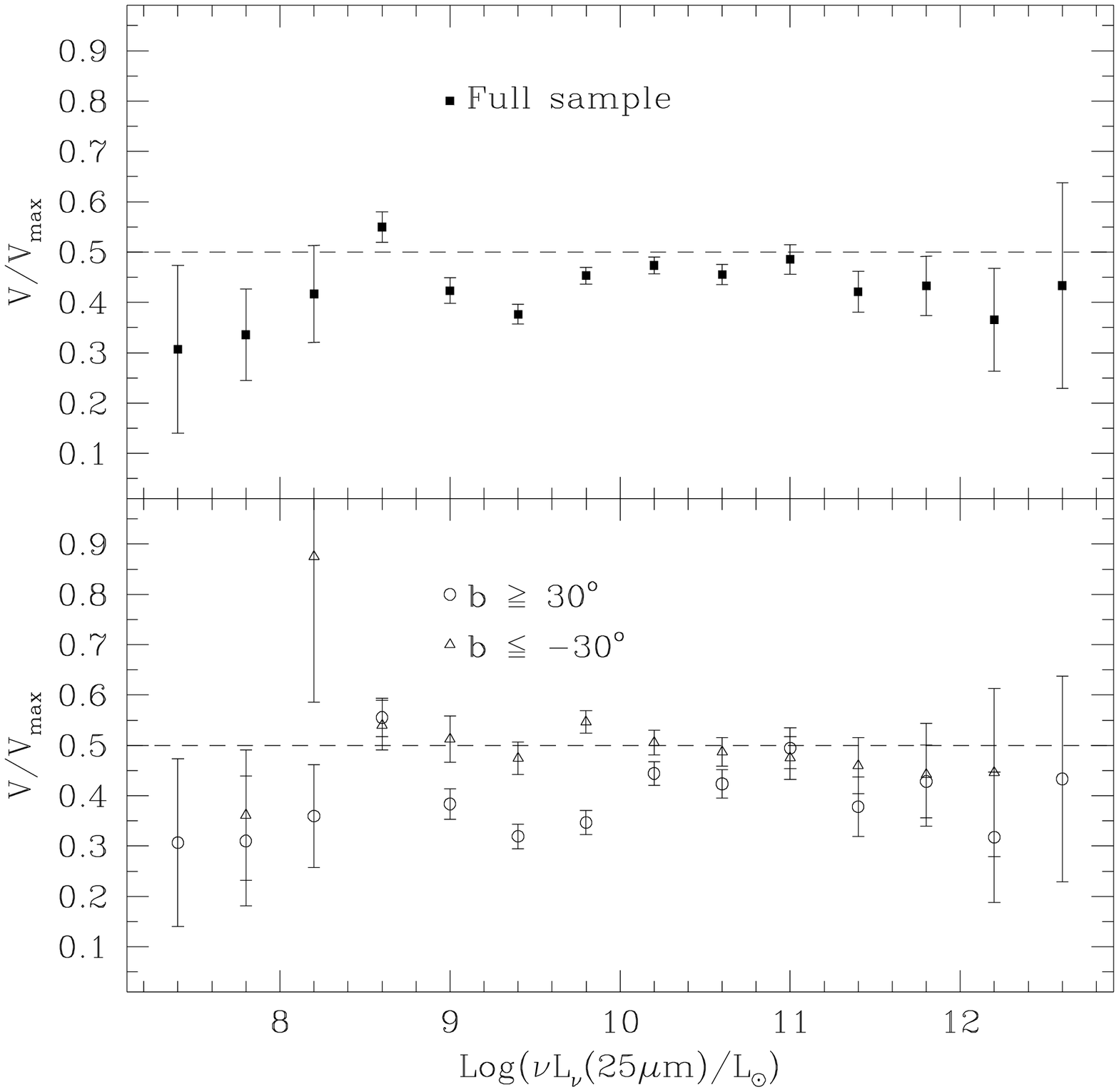]{
$V/V_{max}$ for the full sample (upper panel), and separately for northern and
southern Galactic latitudes (lower panel).  
The dashed line at $V/V_{max}=0.5$ shows the
expected value for a uniform distribution of galaxies. 
\label{fig:vratio1}}

\figcaption[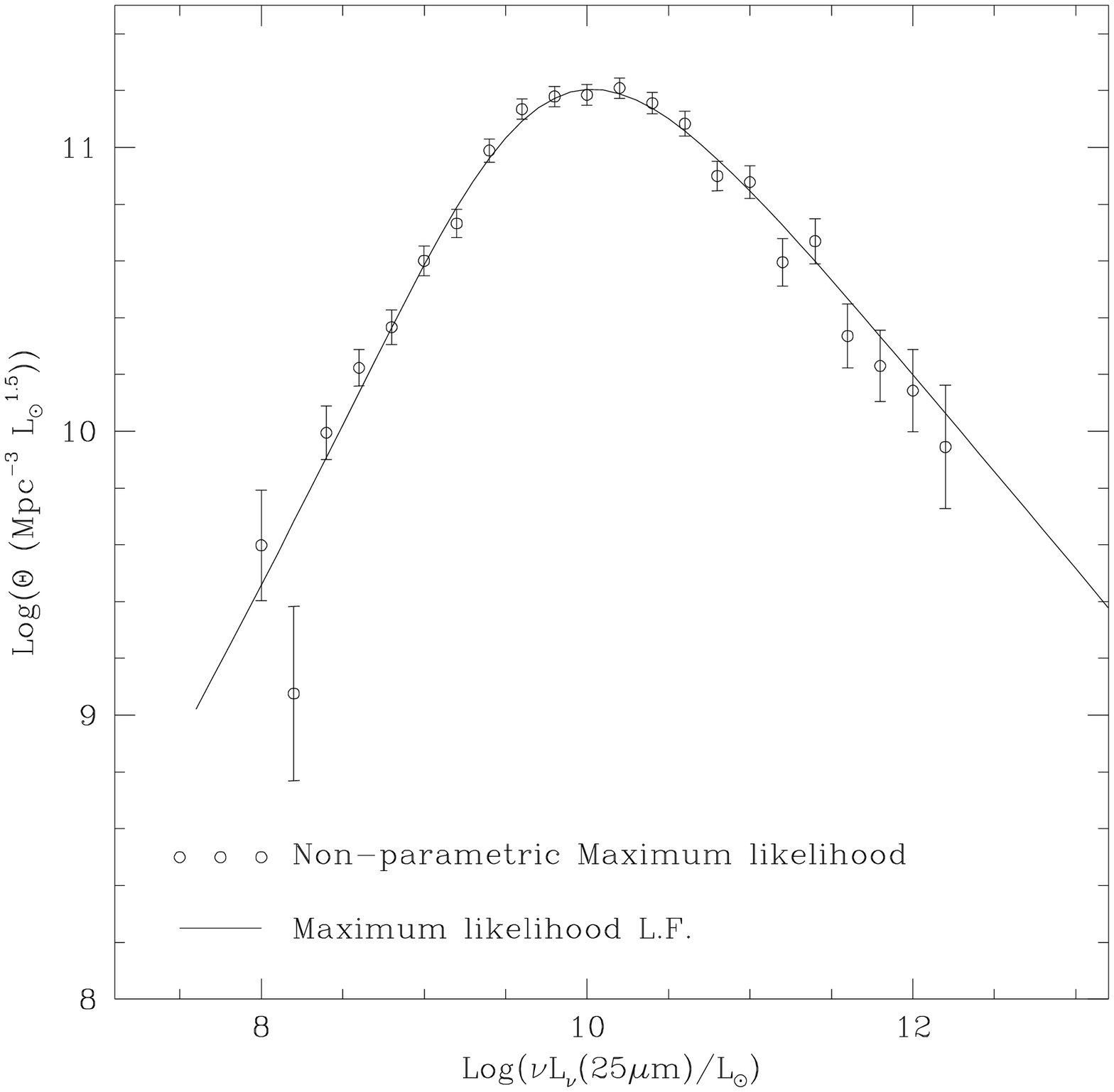]{ 
The non-parametric maximum-likelihood estimate
of the visibility function (circles).   The parametric maximum-likelihood estimate from
Figure \ref{fig:newvis} is also shown for comparison.  
The different estimators agree within the $1\sigma$ error bars shown on
the Figure.
\label{fig:lf_swml}}

\figcaption[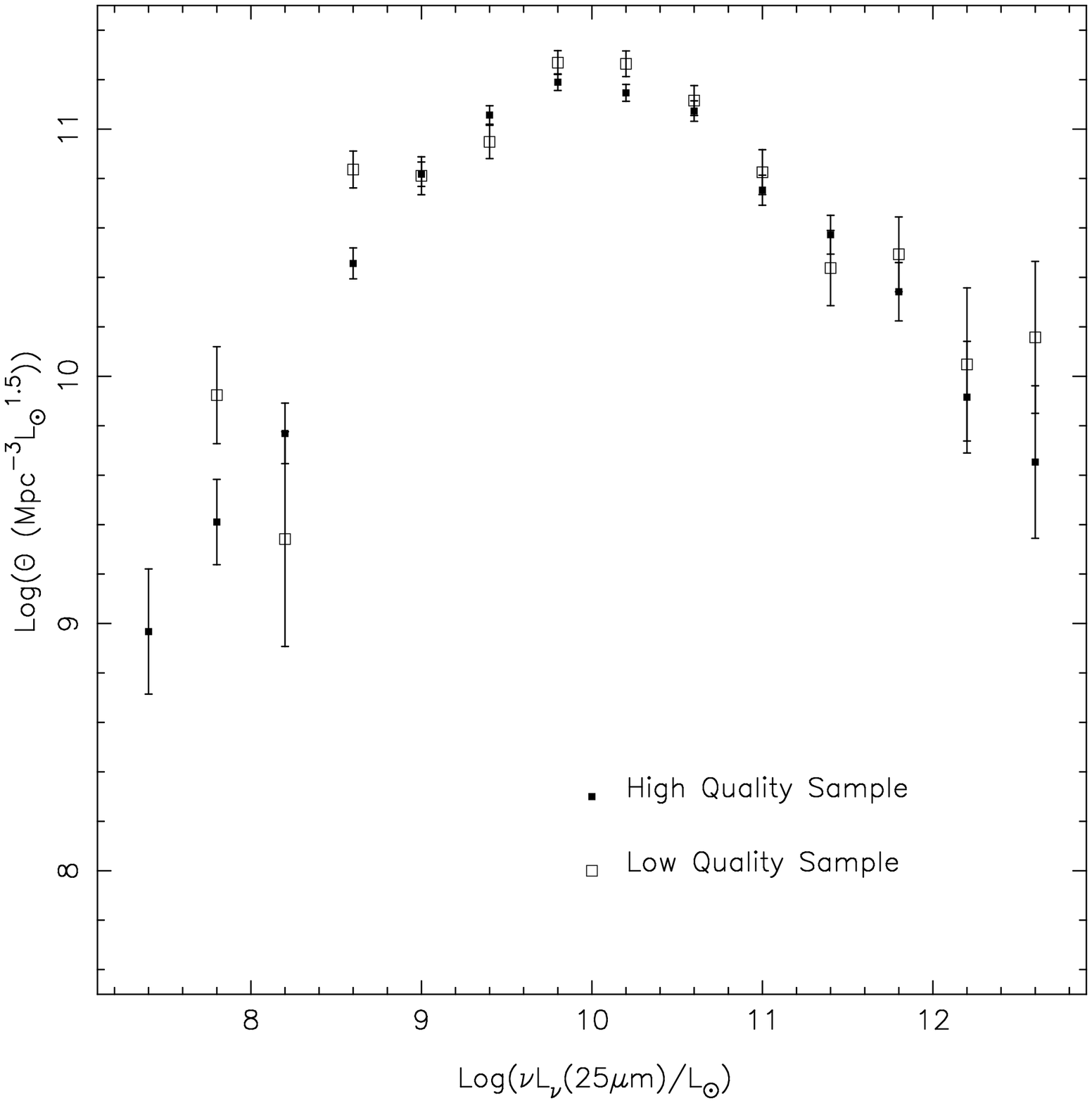]{
The $1/V_{max}$ visibility functions for the high-quality (open squares) and
low-quality subsamples (filled squares) defined in Section \ref{sec:compdisc}.
The $1\sigma$ error bars are shown.
\label{fig:comphighlowQ}}

\figcaption[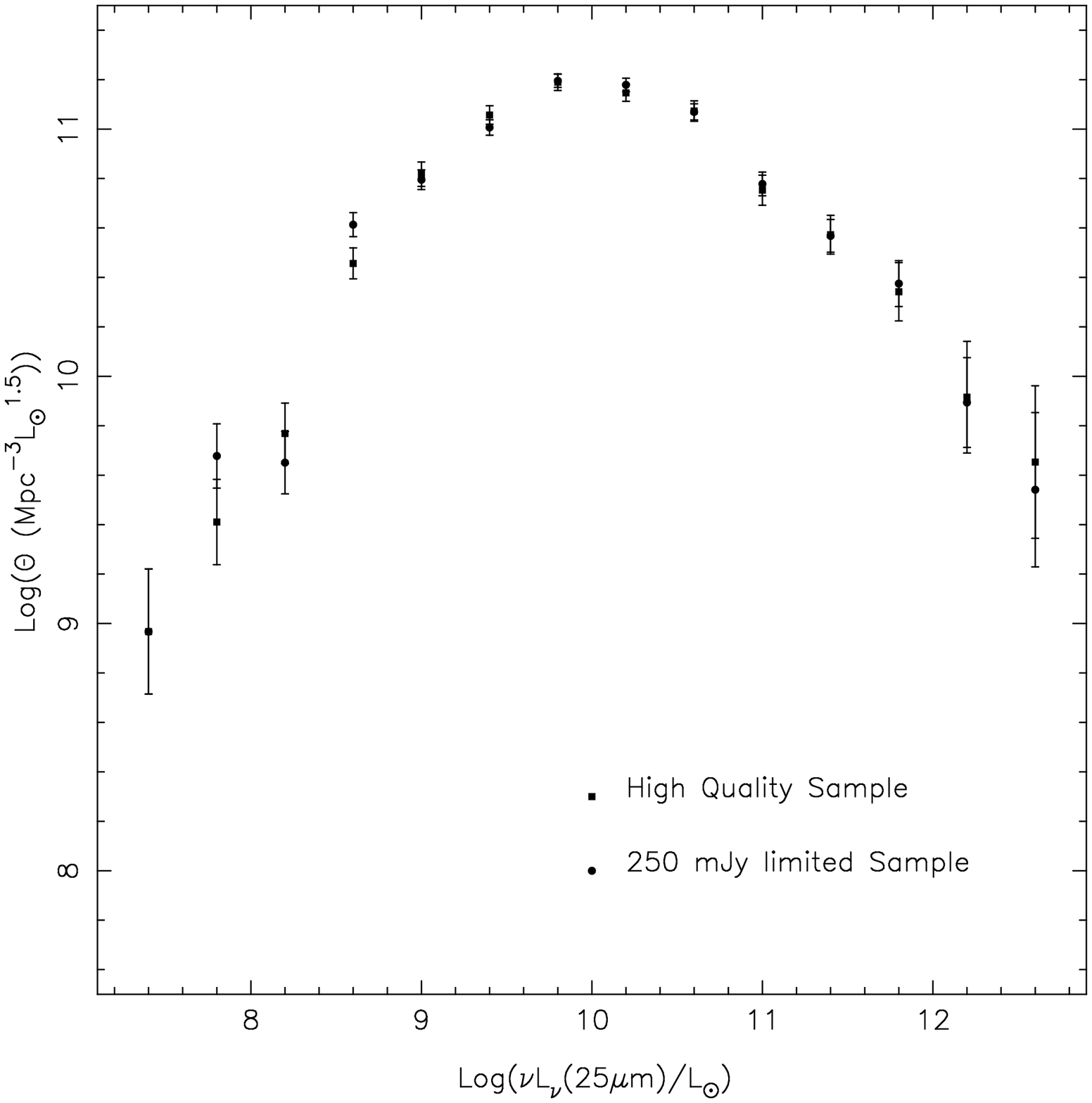]{
The $1/V_{max}$ visibility functions for the high-quality subsample (filled-squares)
defined in Section \ref{sec:compdisc} and the full 250 mJy-limited
sample (filled-circles).  Also shown are the $1\sigma$ error bars.
\label{fig:comphighQwhole}}

\figcaption[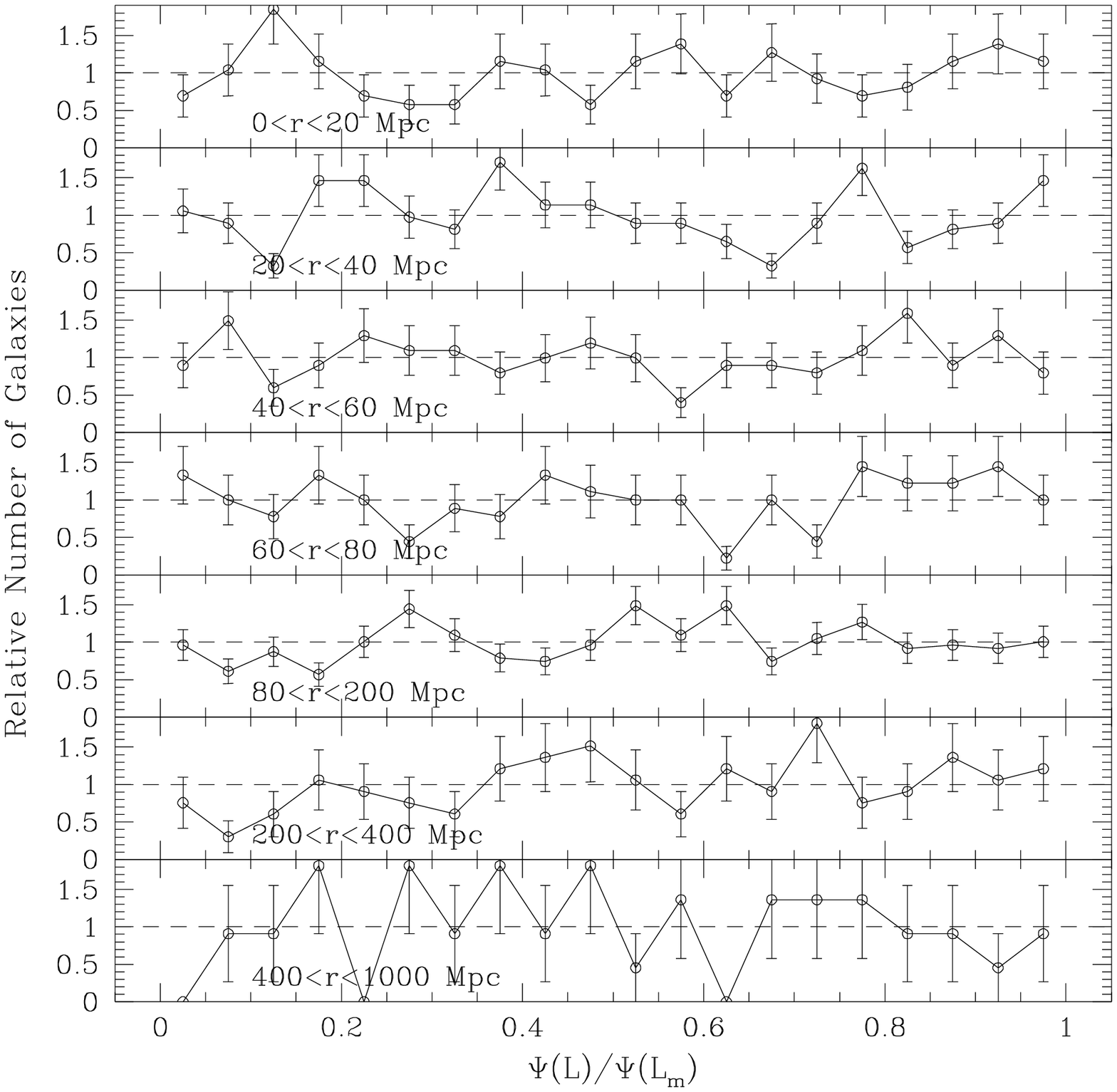]{
Distribution of the quantity $\Psi(L)/\Psi(L_{min})$ in
the sample for galaxies in different distance ranges.  The mean
value is normalized to 1 in each distance range.  The flatness of
these distributions confirms that the completeness of the
sample is high to at
least 400 Mpc.
\label{fig:compcheck}}

\figcaption[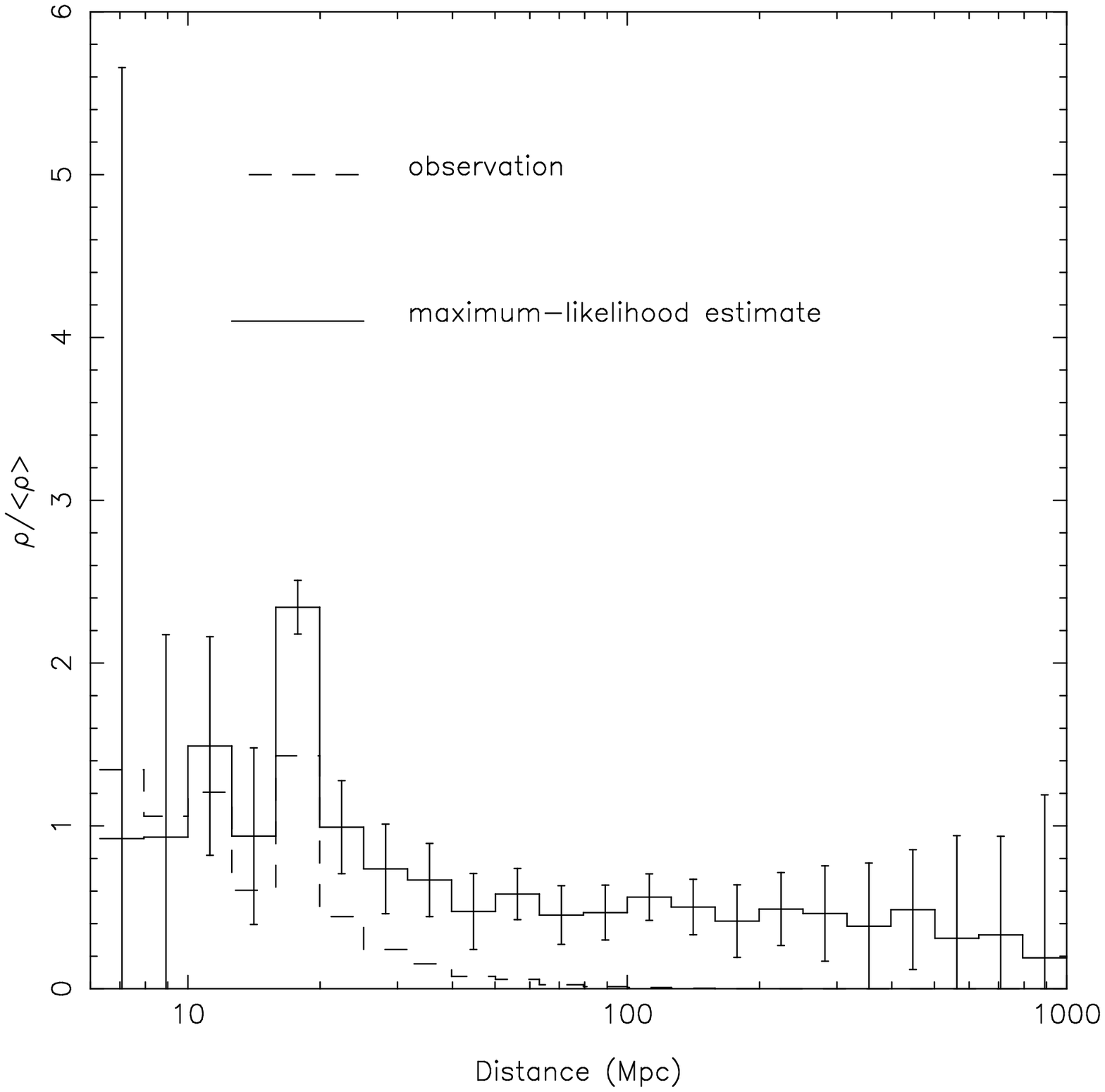]{ 
The non-parametric maximum-likelihood
estimate (solid histogram) of the radial density distribution for the
250 mJy flux-limited sample at 25 $\mu$m.  This estimate is independent
of the galaxy luminosity function.  The dashed lines show the observed
radial density distribution.
\label{fig:den_swml}}

\figcaption[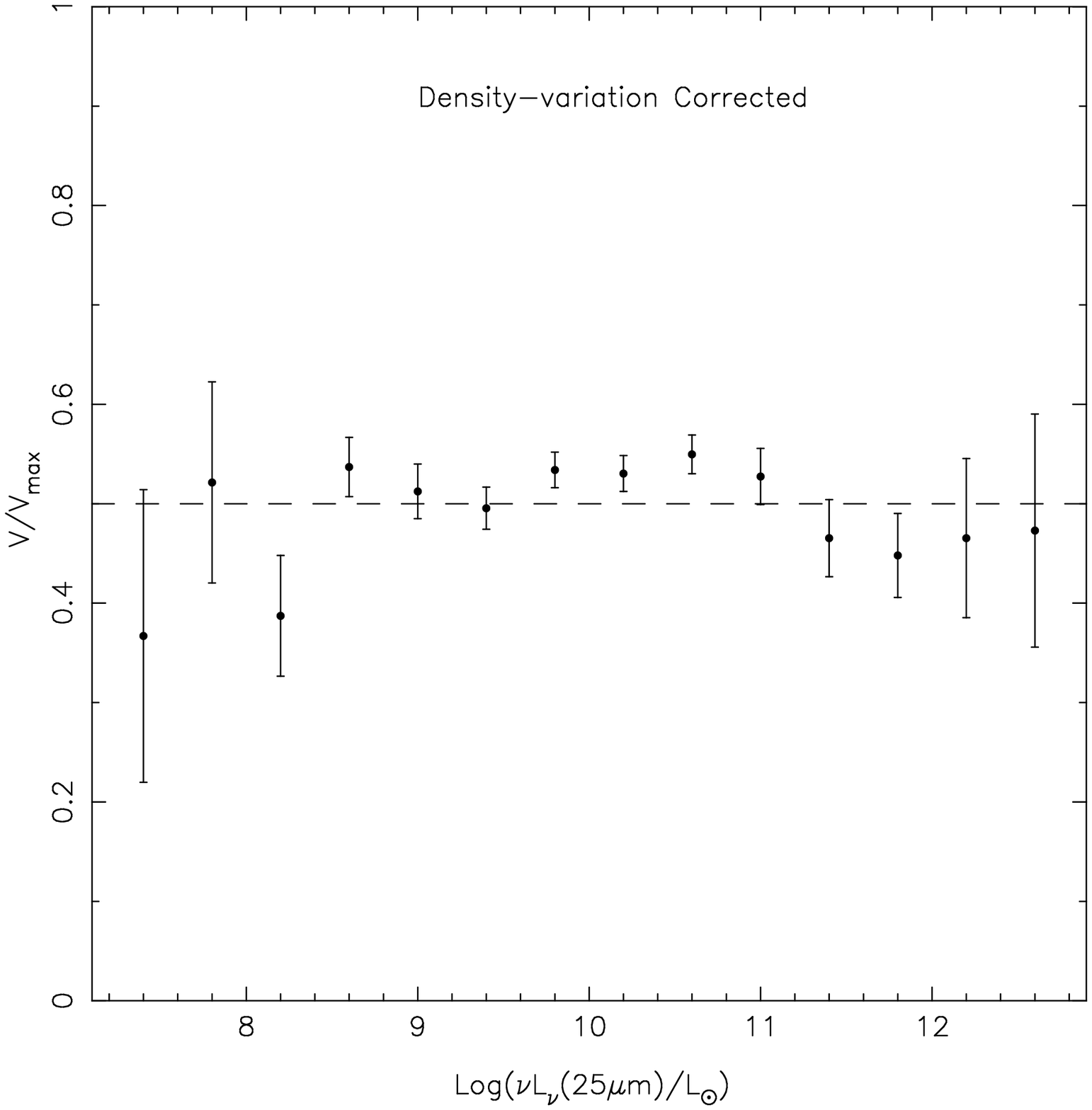]{
The $V/V_{max}$ plot for the full flux-limited sample after correcting for
density variations as described in the text.  The
$1\sigma$ error bars are shown.  
\label{fig:vmax2}}

\figcaption[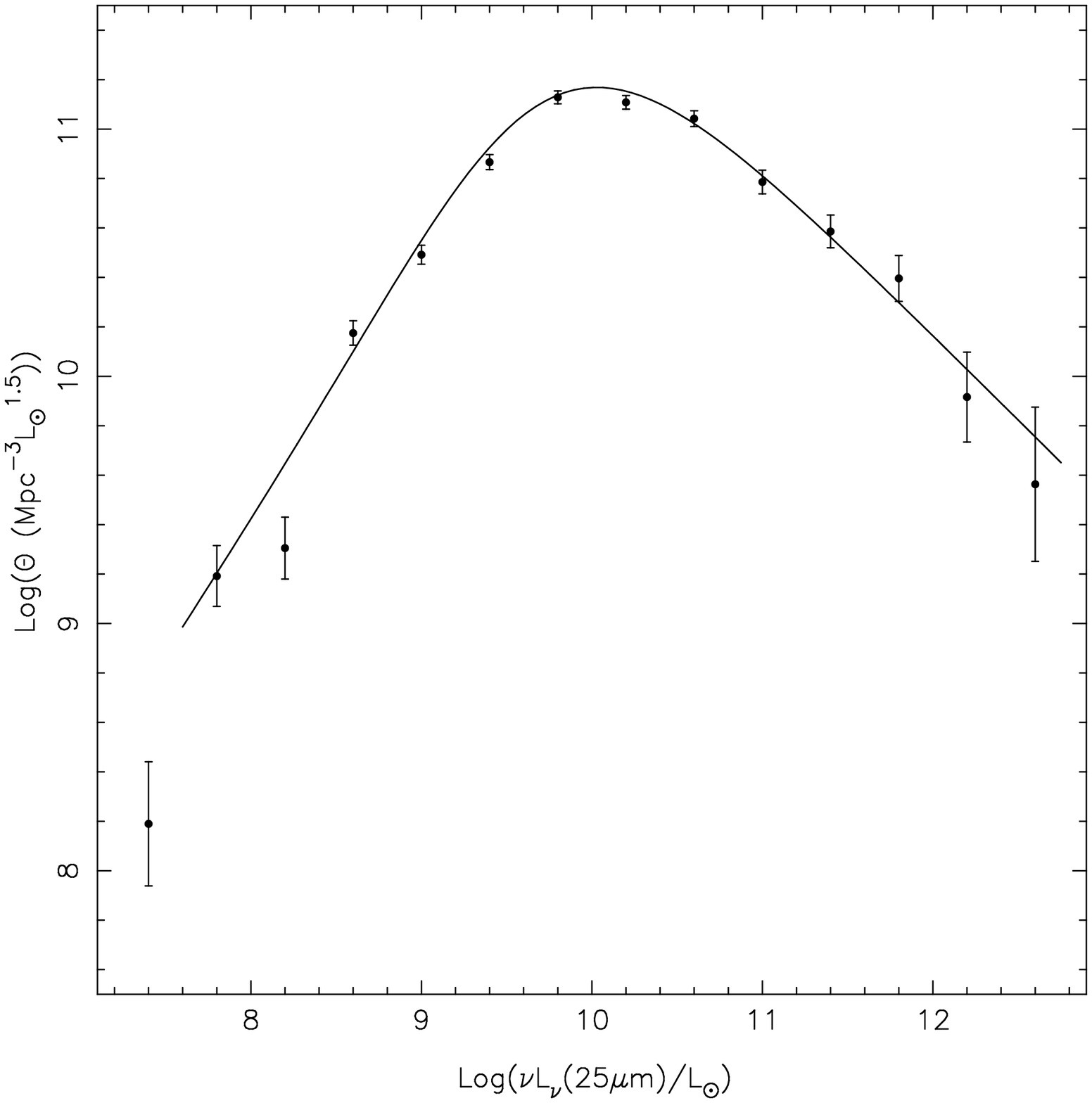]{ 
The density-corrected $1/V_{max}$ estimate
of the visibility function.    The parametric maximum-likelihood estimate from
Figure \ref{fig:newvis} is also shown for comparison.  The density 
inhomogeneities have been
corrected according to the radial density estimate shown 
in Figure \ref{fig:den_swml}.
The different estimators agree within the $1\sigma$ error bars shown on
the Figure.
\label{fig:lfvmax}}

\figcaption[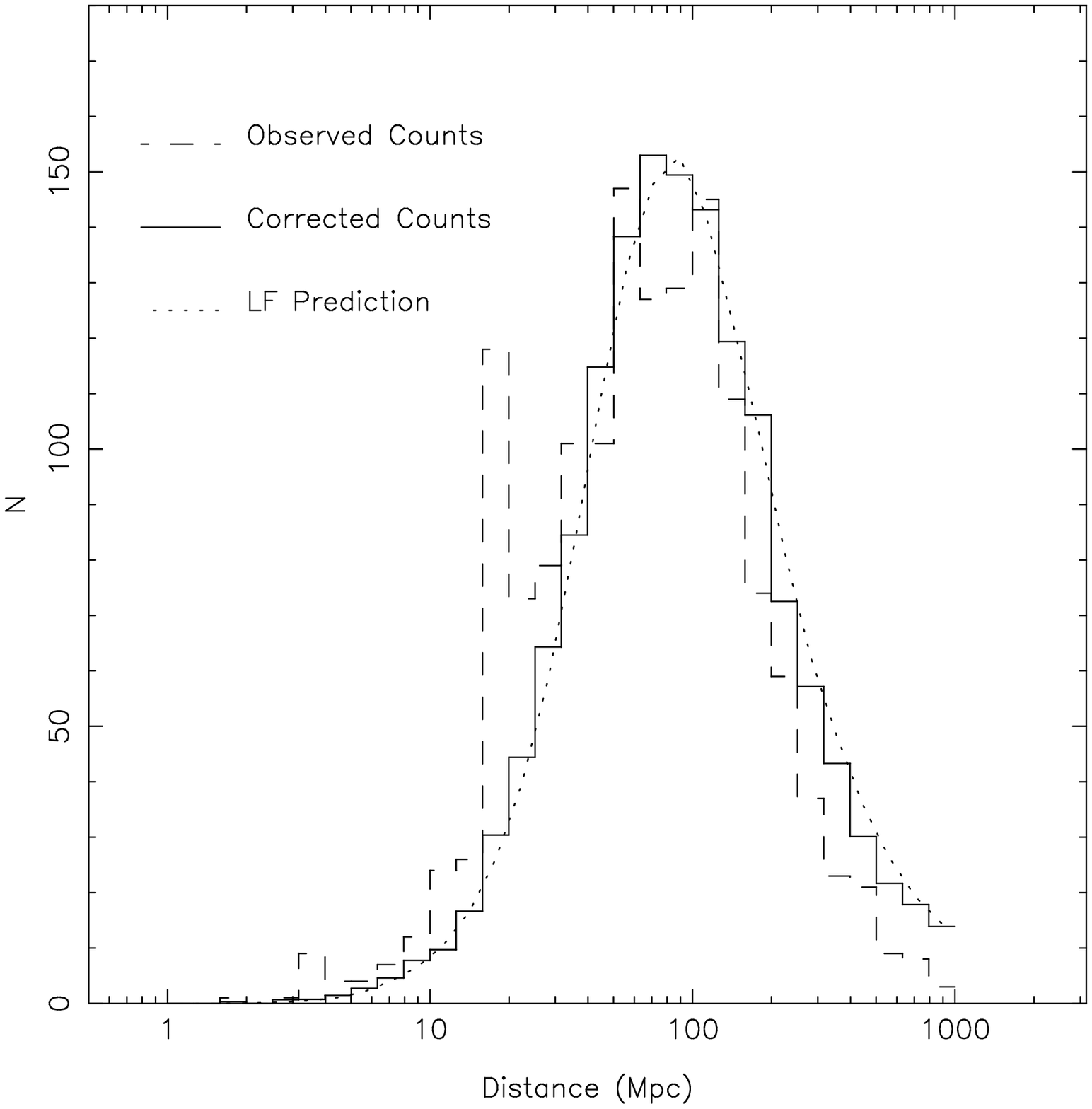]{
The observed redshift distribution (dashed-line histogram) and the
distribution corrected for the density variations in Figure
\ref{fig:den_swml} (solid-line histogram) as described in the text.  The
dotted line is the distribution predicted by the parametric
maximum-likelihood luminosity function from
Figure \ref{fig:newvis} and Table \ref{tab:fitpars}.  \label{fig:srccnt}}

\figcaption[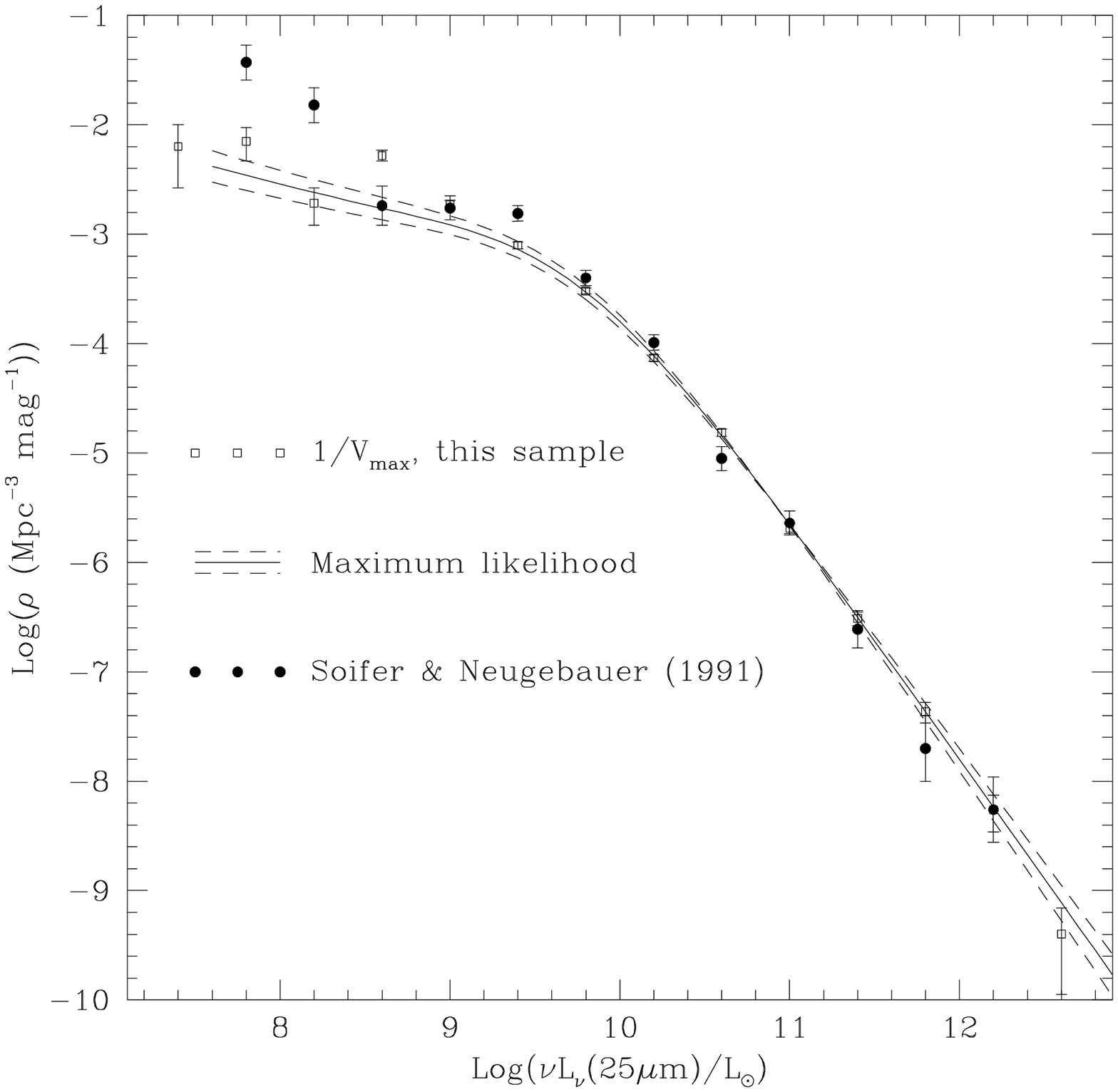]{
Local luminosity functions at 25 $\mu$m plotted in terms of
space density as a function of luminosity.  The dashed lines were
computed by varying the values in Table \ref{tab:fitpars} by their
$1\sigma$ uncertainties.
\label{fig:spacedens}}

\figcaption[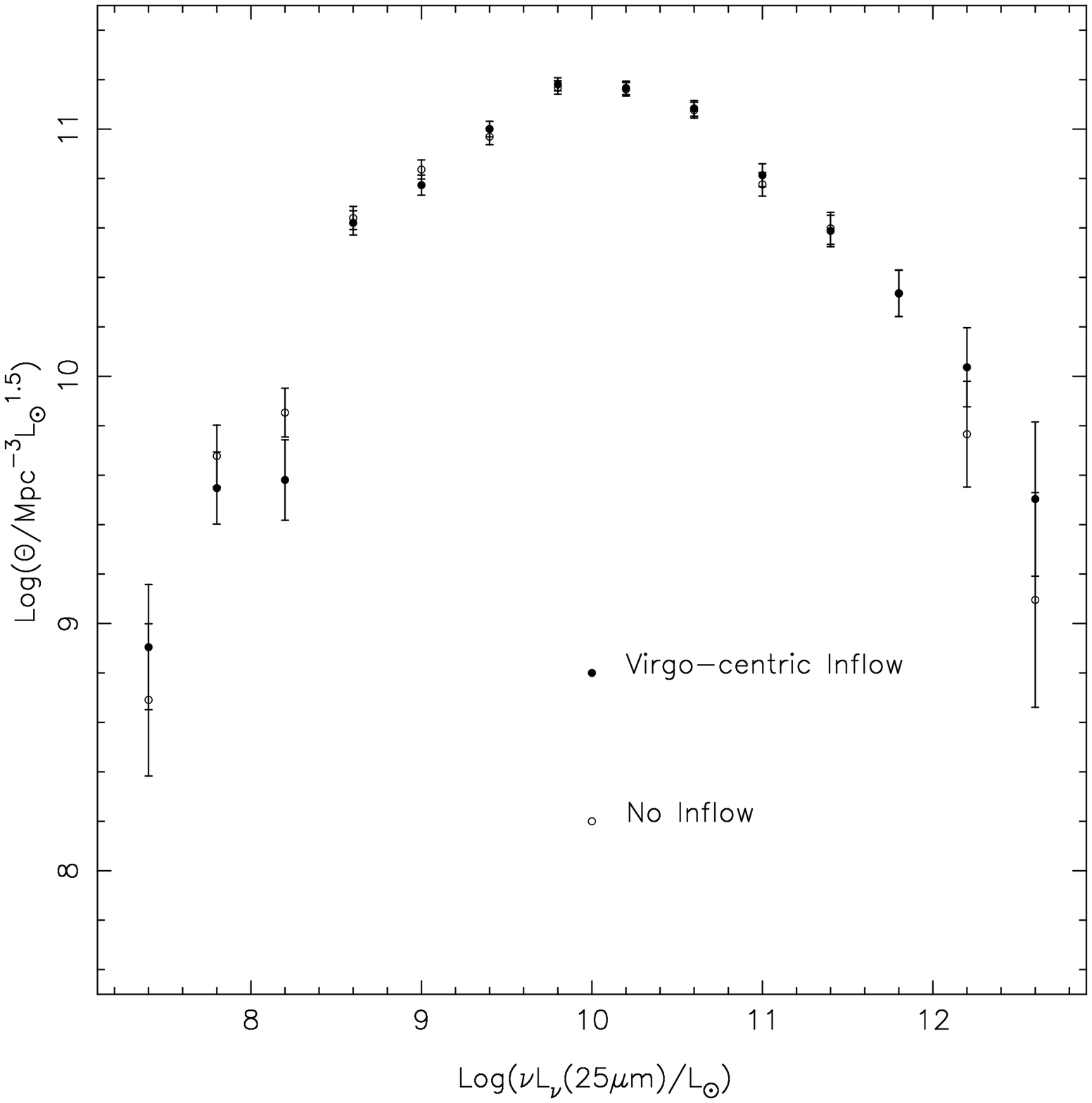]{
The $1/V_{max}$ visibility functions for the 250 mJy-limited sample
with (filled circles, as in Figure \ref{fig:comphighQwhole}) and without
(circles) the Virgo-centric inflow.  The two functions agree well
within the $1\sigma$ error bars, indicating that
most of the galaxies in the sample are not affected by the inflow model.
\label{fig:flowtst}}

\clearpage

\begin{deluxetable}{lr}
\tablewidth{0pt}
\tablecaption{Source of redshifts for sample\label{tab:redsrc}}
\tablehead{
\colhead{Source of redshift}   &  \colhead{Number}}
\startdata
Public Sources: & \nl
1.2 Jy catalog       &   1277  \nl
Nov. 1993 ZCAT       &     49  \nl
NED                  &     79  \nl
SIMBAD               &      6  \nl
\markcite{fair88}Fairall, Lowe, \& Dobie (1988) & 1 \nl
\nl
Private Sources: & \nl
FLWO 1.5m (Table \ref{tab:redshifts})     &     41  \nl
QDOT (\markcite{law94}Lawrence et al.\ 1994) & 5 \nl
\enddata
\end{deluxetable}

\clearpage

\begin{deluxetable}{llcc}
\tablewidth{0pt}
\tablecaption{Previously unpublished redshifts from FLWO\label{tab:redshifts}}
\tablehead{
\colhead{R.A.}   &  \colhead{Dec.} & \colhead{$F_\nu(25)$} & \colhead{c$z$} \nl
\multicolumn{2}{c}{(B1950)} & (Jy) & (km s$^{-1}$)}
\startdata
00 17 02.7  & $-$04 56 46  &   0.3734  &  $ 6181 \pm 28$   \nl  
00 20 47.6  & +10 29 50    &   0.2633  &  $69094 \pm 85$  \nl  
00 37 17.8  & +08 29 18    &   0.2619  &  $17482 \pm 34$   \nl  
00 44 30.0  & $-$18 03 33  &   0.2776  &  $43979 \pm 48$  \nl  
00 44 37.2  & +10 14 46    &   0.3265  &  $\begin{array}{c}49959 \pm 54 (N) \\
                                   50133 \pm 43 (S) \end{array}$    \nl  
00 56 04.2  & +08 31 54    &   0.2861  & $ 17269 \pm 36$  \nl  
01 27 44.3  & +07 53 04    &   0.2541  &  $33899 \pm 44$ \nl  
01 47 43.0  & $-$16 55 12  &   0.2875  &  $48432 \pm 39$ \nl  
07 55 56.0  & +50 58 36    &   0.3589  &  $16305 \pm 41$ \nl  
08 11 42.0  & +46 13 00    &   0.2951  &  $12277 \pm 136$  \nl  
08 46 21.1  & +11 26 03    &   0.2626  &  $23218 \pm 33$    \nl  
08 57 06.0  & +18 01 00    &   0.2645  &  $\begin{array}{c}15940 \pm 20 (E) \\
                                   15830 \pm 19 (W) \end{array}$   \nl  
09 39 16.1  & +23 54 38    &   0.3299  &  $ 6407 \pm 39$  \nl  
09 42 35.4  & +17 51 45    &   0.4363  &  $38440 \pm 41$    \nl  
10 03 12.0  & +13 12 00    &   0.4967  &  $2789 \pm 28$   \nl  
10 07 43.0  & +14 27 39    &   0.2729  &  $59689 \pm 55$   \nl  
10 08 05.4  & +06 26 45    &   0.2883  &  $29445 \pm 39$    \nl  
10 23 48.0  & +00 56 00    &   0.3076  &  $ 6422 \pm 18$  \tablebreak  
11 06 26.0  & $-$23 21 57  &   0.3648  &  $24551 \pm 41$  \nl  
11 19 58.7  & +04 31 06    &   0.5010  &  $11399 \pm 38$  \nl 
11 22 24.9  & +02 56 02    &   0.2533  &  $\begin{array}{c}18813 \pm 17 (N) \\
                                   18857 \pm 24 (S) \end{array}$ \nl 
11 31 56.3  & +22 47 49    &   0.3011  &   $9230 \pm 34$  \nl  
11 57 29.2  & $-$03 30 20  &   0.2663  &   $8024 \pm 20$   \nl  
12 12 17.0  & $-$03 12 00  &   0.4950  &  $10108 \pm 37$    \nl  
12 17 18.6  & +09 12 59    &   0.2743  &  $ 7398 \pm 31$   \nl  
12 19 41.7  & +00 09 03    &   0.2991  &   $\begin{array}{c}51652 \pm 29 (N) \\
                                   51626 \pm 36 (S) \end{array}$     \nl  
12 23 25.0  & $-$23 19 36  &   0.3232  &  $14568 \pm 40$   \nl  
12 36 16.3  & $-$27 02 00  &   0.2678  &  $ 7502 \pm 22$    \nl  
12 38 18.7  & +27 50 19    &   0.3110  &  $16947 \pm 34$    \nl  
12 50 30.4  & $-$02 56 36  &   0.4896  &  $ 6807 \pm 34$   \nl  
13 24 15.4  & $-$06 08 38  &   0.2963  &  $13489 \pm 46$    \nl  
13 53 28.3  & $-$07 51 36  &   0.3494  &  $22750 \pm 40$  \nl  
13 55 57.1  & $-$15 53 51  &   0.3377  &  $10868 \pm 34$  \nl  
15 19 54.4  & $-$13 51 29  &   0.2555  &  $ 7182 \pm 25$   \nl  
15 21 47.0  & $-$11 02 07  &   0.2667  &  $15447 \pm 37$   \nl  
15 21 48.0  & +05 01 00    &   0.2588  &  $10715 \pm 10$  \tablebreak  
21 11 45.0  & $-$27 31 29  &   0.2655  &  $25801 \pm 33$   \nl  
22 34 59.5  & $-$25 44 16  &   0.2646  &  $23353 \pm 21$    \nl  
23 13 02.6  & $-$00 58 50  &   0.3414  &  $ 8318 \pm 36$   \nl  
23 29 00.6  & $-$12 52 29  &   0.3778  &  $ 6357 \pm 40$  \nl  
23 40 45.2  & +04 16 37    &   0.3489  &  $\begin{array}{c}11478 \pm 23 (W) \\
                                   11564 \pm 21 (N) \\
                                   11334 \pm 21 (S) \end{array}$  \nl  
\enddata
\tablecomments{Multiple redshift components are labeled with letters
denoting their relative orientations.}
\end{deluxetable}

\clearpage

\begin{deluxetable}{ccccc}
\tablewidth{0pt}
\tablecaption{Fitted Parameters of Luminosity Function\label{tab:fitpars}}
\tablehead{\colhead{$\alpha$} & \colhead{$\beta$} & 
\colhead{$L_*$ } & \colhead{$C$} & \colhead{$n_1$} \nl
~ & ~ & \colhead{($L_\odot$)} & \colhead{($10^{-3}$ Mpc$^{-3}$)} & 
\colhead{($10^{-3}$ Mpc$^{-3}$)} 
}
\startdata
$0.437 \pm 0.032$ & $1.749 \pm 0.067$ & $4.07 \pm 0.36 \times 10^9$ &
$1.23 \pm 0.09$ & $9.2 \pm 0.7$ \nl
\enddata
\end{deluxetable}

\end{document}